# Experimental Study of a Low-Voltage Pulsed Plasma Thruster for Nanosatellites

Patrick Gresham[1], Brian Jeffers[2], Alexey Shashurin[3]

*School of Aeronautics and Astronautics, Purdue University, West Lafayette, IN 47907, USA*

**In this paper, a coaxial pulsed plasma thruster (PPT) was designed, built, and tested. This design confirmed PPT operation at substantially reduced discharge voltages of $100 - 200\ V$ across the discharge plasma at a discharge current level of $10.4\ kA$ similar to magnetoplasmadynamic arcs. The PPT cathode was imaged with an ICCD camera over a wide range of pressures, and the photos indicated "spotless" diffuse arc attachment to the cathode at higher pressures, and the appearance of conventional cathode spots at lower pressures. Cathode erosion rate measurements support the conclusion of "spotless" operation at higher pressures and the presence of cathode spots at lower pressures (the erosion rates of 16.194 µg/C and 40.785 µg/C were measured at 4 Torr and 3.5×10$^{-5}$ Torr, respectively).**

## I. Nomenclature

| | | |
|---|---|---|
| *g* | = | *grams* |
| *C* | = | *Coulombs (Charge)* |
| *T* | = | *Thrust [N]* |
| *s* | = | *seconds* |
| *min.* | = | *minutes* |
| *m* | = | *meters* |
| *V* | = | *Volts (Voltage)* |
| *A* | = | *Amperes (Current)* |
| *B* | = | *Magnetic Field [T]* |
| *j* | = | *Current Density [A/m$^2$]* |
| *Hz* | = | *Hertz [s$^{-1}$]* |
| *mgs* | = | *milligrams per second [mg/s]* |
| *sccm* | = | *Standard Cubic Centimeters per Minute [cm$^3$/min.]* |
| *PPT* | = | *Pulsed Plasma Thruster* |
| *MPD* | = | *Magnetoplasmadynamic* |
| *LESF* | = | *Low Energy Surface Flashover* |
| *EPPL* | = | *Electric Propulsion and Plasma Laboratory* |
| *PTFE* | = | *Polytetrafluoroethylene* |
| *ICCD* | = | *Intensified Charge Coupled Device* |

## II. Introduction

Pulsed Plasma Thrusters (PPT) have many characteristics that make them well suited for nanopropulsion applications. They are small, have low power requirements, and have shown good reliability [1]. They provide precise impulse bits and maintain constant specific impulse over a wide range of input power levels [2]. Additionally, PPT's are quite flexible, as many different types of geometries, energy storage capacitors, and feed/ignition systems can be used [3], and they can easily be recharged from solar panels [4].

---

[1] Equal Authorship, Masters Alumnus, School of Aeronautics and Astronautics
[2] Equal Authorship, Masters Student, School of Aeronautics and Astronautics
[3] Associate Professor, School of Aeronautics and Astronautics

Historically, most satellite PPT designs are ablative in nature, and often use polytetrafluoroethylene (PTFE) or a similar fluorocarbon as a solid-phase propellant; these solid-phase PPT's have a long flight heritage as high specific impulse micropropulsion systems [5,6]. However, PTFE PPT's used in orbit have achieved efficiencies of only 5 to 10% [5], and they are plagued by issues such as molten macroparticles, propellant charring, and nonuniform propellant ablation rates [7], as well as carbon contamination to spacecraft solar panels [8]. There are two primary factors that result in a rather limited lifetime for these systems, those being changes to the propellant surface topology [9], and igniter failure after extended operation [10].

Research into gas-fed PPT's (GF-PPT's) has been performed to overcome some of these challenges. GF-PPT's have high efficiencies [11][12], but at the cost of high-pressure tanks and complex gas injection systems which strain the volume limitations present on the CubeSat platform. While the problem of nonuniform ablation was eliminated, there was the new difficulty of timing the gas injection and ignition properly, and GF-PPT's are yet to be used on nanosatellites [9].

The remaining option, which is to use a liquid propellant, has attracted recent attention [13]. Several research projects have been conducted on LF-PPT's, and a review conducted by Rezaeiha and Schoenherr found that liquid propellants are promising for PPT performance optimization [13]. LF-PPT's were found to have higher efficiencies and specific impulses than solid-phase PPT's, and do not suffer from nonuniform propellant ablation, but without needing to carry high-pressure propellant tanks like GF-PPT's [13]. Additionally, successful operation has been demonstrated with low contamination risk liquids such as water [13]. As a result, a liquid-fed version of a pulsed plasma thruster might offer performance advantages over legacy solid-phase versions, while also solving the problem of long operational lifetime.

Recently, we developed and studied a liquid-fed pulsed plasma accelerator using a Low Energy Surface Flashover (LESF) assembly as the igniter [4]. Pentaphenyl trimethyl trisiloxane, a diffusion pump oil, was used as the propellant for the PPT's demonstration because its low vapor pressure made it suitable as a test fluid. With the capacitors held at 1.8 kV, a discharge produced 7.42 kA of current and led to an estimated ion exhaust velocity of $32 \pm 4 \frac{km}{s}$, which corresponds to a thrust value on the order of $T \approx 5.8 \pm 0.7\ N$ and an impulse bit of $35 \pm 4.4\ \mu N \cdot s$ [4], which demonstrates that an LF-PPT with an LESF igniter can deliver performance in the expected range of a standard PPT. Subsequently, AF-M315E was loaded as the propellant, and there was a successful ignition and discharge, demonstrating its capability to function as an LF-PPT propellant [4].

This initial work by Patel et al. had two disadvantages for practical applications. Firstly, the capacitors being held at nearly 2 kV would require step-up converters on-board as nanosatellite power supplies typically only provide low voltages. These converters would add additional volume and weight to the propulsion system, which is a large downside considering it is essential to make nanosatellite systems as compact and light as possible. Additionally, utilization of a step-up converter and high operational voltages (several kV) create new electrical concerns, such as unwanted sparking between the system elements, and require thorough electrical insulation which is also associated with added system weight and volume. Secondly, the long discharge electrodes utilized (6.5 cm in Ref. [4]) resulted in energy losses as the discharge pulse developed. Specifically, the discharge current passed through an increasing length of the electrodes as the plasma layer accelerated and propagated along the thruster electrodes. As high discharge currents were utilized (~7 kA), substantial voltage drops along the discharge electrodes are expected, causing ohmic losses in the electrodes and taking away from the kinetic energy spent on the plasma layer acceleration.

This work pursued the goals of reducing the operational voltage and preventing energy losses in the discharge electrodes discussed above. First, we aimed to create a highly conductive interelectrode plasma channel associated with low voltage drops similar to that in magnetoplasmadynamic (MPD) arcs (<100 V). Second, we redesigned the PPT geometry and removed long discharge electrodes to eliminate the corresponding voltage drops and energy losses.

### III. Methodologies & Equipment

Compared to the parallel plate PPT from Patel et al., this design of the coaxial PPT used decreased lengths of the anode and cathode electrodes to further decrease corresponding voltage drops. The system consisted of a thoriated tungsten cathode rod (sharpened at the end facing the discharge) and a copper ring anode inside a clear KF-25 glass nipple, as shown in Figure 1. This setup allowed for high quality images of the cathode.

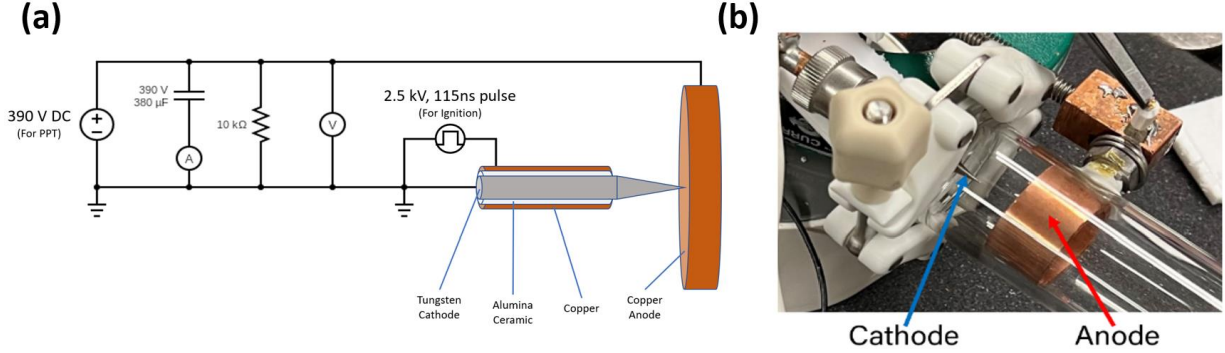

*Figure 1: (a) Circuit diagram and (b) photograph of the coaxial PPT used in this work*

The experiments were conducted in a vacuum chamber evacuated by mechanical and diffusion pumps, enabling residual pressures to reach below $10^{-5}$ Torr. For reading pressures above $10^{-3}$ Torr, a PDR900 vacuum gauge controller with a 925 micropirani transducer by MKS Instruments Inc was used, and for reading pressures below $10^{-3}$ Torr, a Granville-Phillips 330 controller with an iridium-filament ion gauge tube was used [14]. Current measurements were conducted with a Person Current Monitor Model 101, which has a maximum peak current allowance of 50 $kA$. Voltage measurements were taken with a Teledyne Lecroy PP023 10:1 voltage probe. To investigate the discharge attachment to the cathode during the pulse, a Princeton Instruments Pi-Max 4 ICCD camera was used, and images were processed in LightField. A Nikon D7500 camera was also used for long exposure imagery of the tungsten cathode and plasma plume.

Measurements of the tungsten cathode erosion rate were conducted using direct measurements of the cathode's mass loss after prolonged PPT operation. Specifically, a precise digital scale providing an accuracy down to $10^{-4}$ g was used to determine the cathode mass loss ($\Delta m$). We conducted several cathode mass loss measurements after a series of 13,000 pulses (1.5 Hz pulse repetition rate, 2.5 hours total duration) in order to ensure a well-detectable mass loss value. Total charge transfer between the discharge electrodes was determined by integrating the current waveform over the entire duration of the series (13,000 pulses) using $Q = \int I dt$. Finally, the cathode erosion rate was determined by using the following equation: $Er = \frac{\Delta m\ [\mu g]}{Q\ [C]}$.

## IV. Results & Discussion

Two long exposure photographs of the coaxial PPT operation are shown in Figure 2 (using air at a pressure of 1 Torr). A corresponding waveform of the discharge parameters is shown in Figure 3. One can see that the anode voltage was $V_a < 200\ V$, and the peak discharge current was $I_{d,peak} = 10.4\ kA$, which is higher than the $7.2\ kA$ reported in our previous work [4]. It should be noted that the actual discharge voltage is expected to be lower than this 200 V value, as measurements were conducted several centimeters away from the location of the discharge attachment to the electrode, and this voltage drop along the electrode was not subtracted.

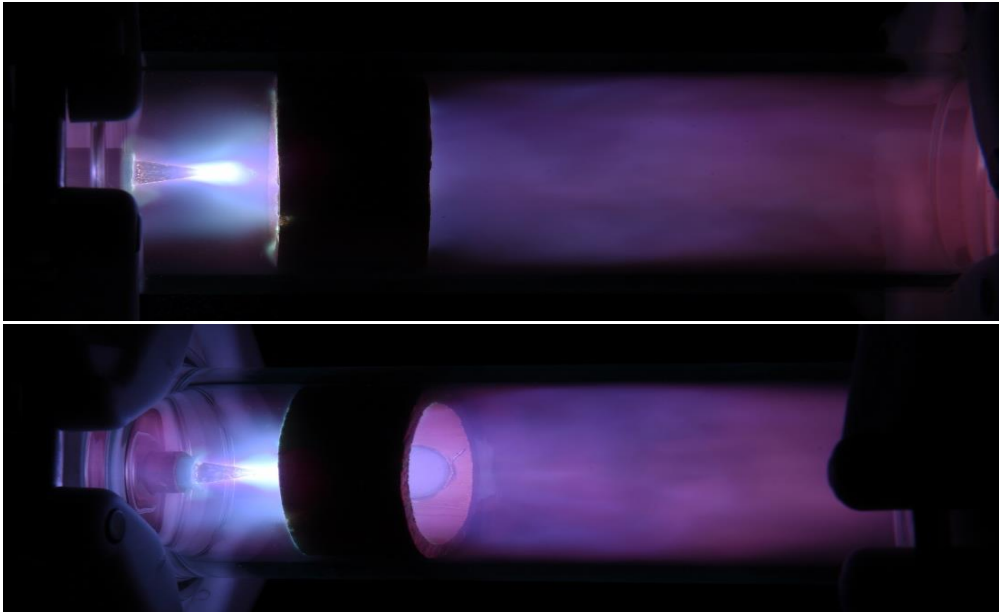

*Figure 2: Long exposure photograph of the coaxial PPT.*

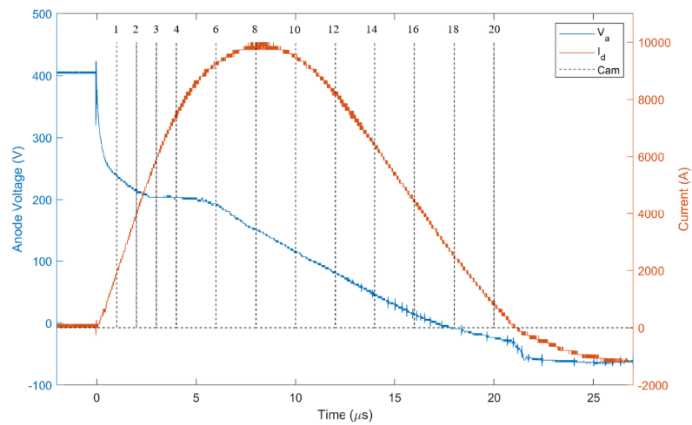

*Figure 3: Image locations relative to discharge current and anode voltage*

A corresponding series of photos were taken with an ICCD camera (using an exposure time of 3 ns) to map out the evolution of the plasma plume and the discharge attachment to the cathode surface at 1 Torr which can be seen in Figure 4 (each image is generated from a separate PPT pulse). The moments in time at which the discharge photos were acquired relative to the electrical waveforms of current and voltage are shown in Figure 3 using dashed vertical lines, and again above the images in Figure 4. The images from Figure 4 reveal that the initial 3 $\mu s$ of the discharge involve the propagation of the plasma layer to the cathode tip, to which the discharge reaches at $t = 3 - 4 \ \mu s$. The remaining dynamics show that the discharge column stays attached to the cathode tip and reaches its peak brightness at $t = 8 - 10 \ \mu s$. The bulk plasma begins to increase in brightness relative to the cathode starting at $t = 14 \ \mu s$. One can clearly see the presence of the volumetric glow of gas around the cathode. Unlike traditional PPT's, this design only has a well-defined plasma sheet for the first 3 $\mu s$, which shows the propagation of the plasma layer. From 3-12 µs, one can see attachment of the discharge column to the tip of the cathode. After 12 µs, a volumetric glow can be seen.

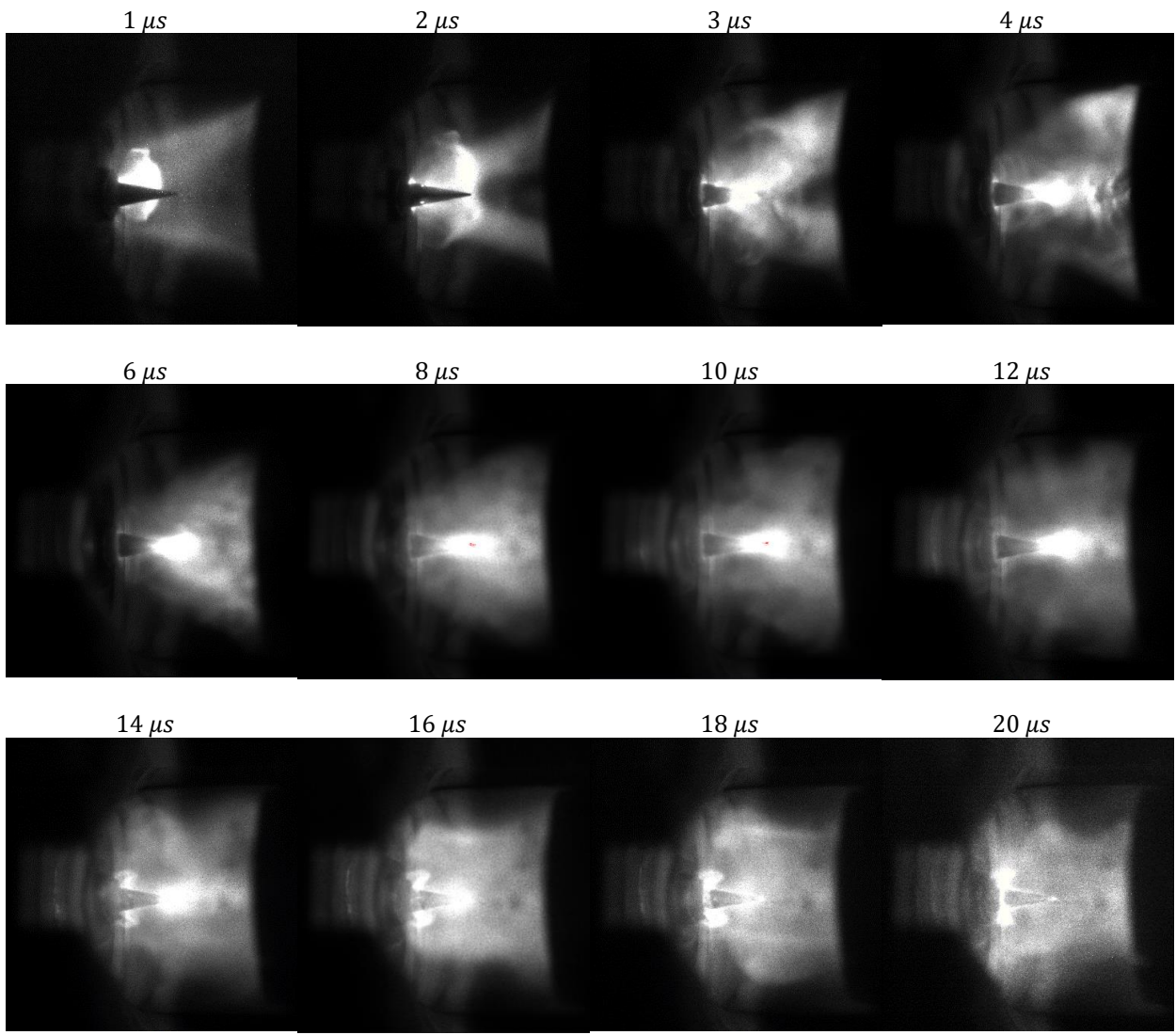

*Figure 4: Plasma Plume Evolution*

Further ICCD photography was conducted for pressures of $10^{-5}, 10^{-4}, 10^{-3}, 10^{-2}, 1,$ and $10$ Torr, as shown in Figure 5. It should be noted that discharge tests at $10^{-1}$ Torr were eliminated from this study, because a DC glow discharge would form between the anode and cathode while attempting to charge the capacitor at that pressure (indicating operation near the Paschen breakdown curve's minimum [15]). The ICCD images are presented for times $t = 3, 8,$ and $14\ \mu s$. All images were taken with a gate width of 3 nanoseconds (the following ICCD scaling was used in LightField: for 3 $\mu s$ - B=600 and W=4,000; for 8 $\mu s$ - B=400 and W=16,000; and for 14 $\mu s$- B=600 and W=4,000). In the ICCD imagery, the discharge can be seen establishing a stationary attachment to the cathode tip shortly after the ignition and creating the glow observed at the cathode tip. Similar "spotless" diffuse attachment of arcs was previously reported when cathode materials with high melting temperature were used [16]. Additionally, large volumetric glow was observed at high pressures (1 and 10 Torr), indicating that the discharge column (conductivity in the gap) is supported substantially by the ionization of the interelectrode gas volume rather than by the metallic plasma originating from cathode spots.

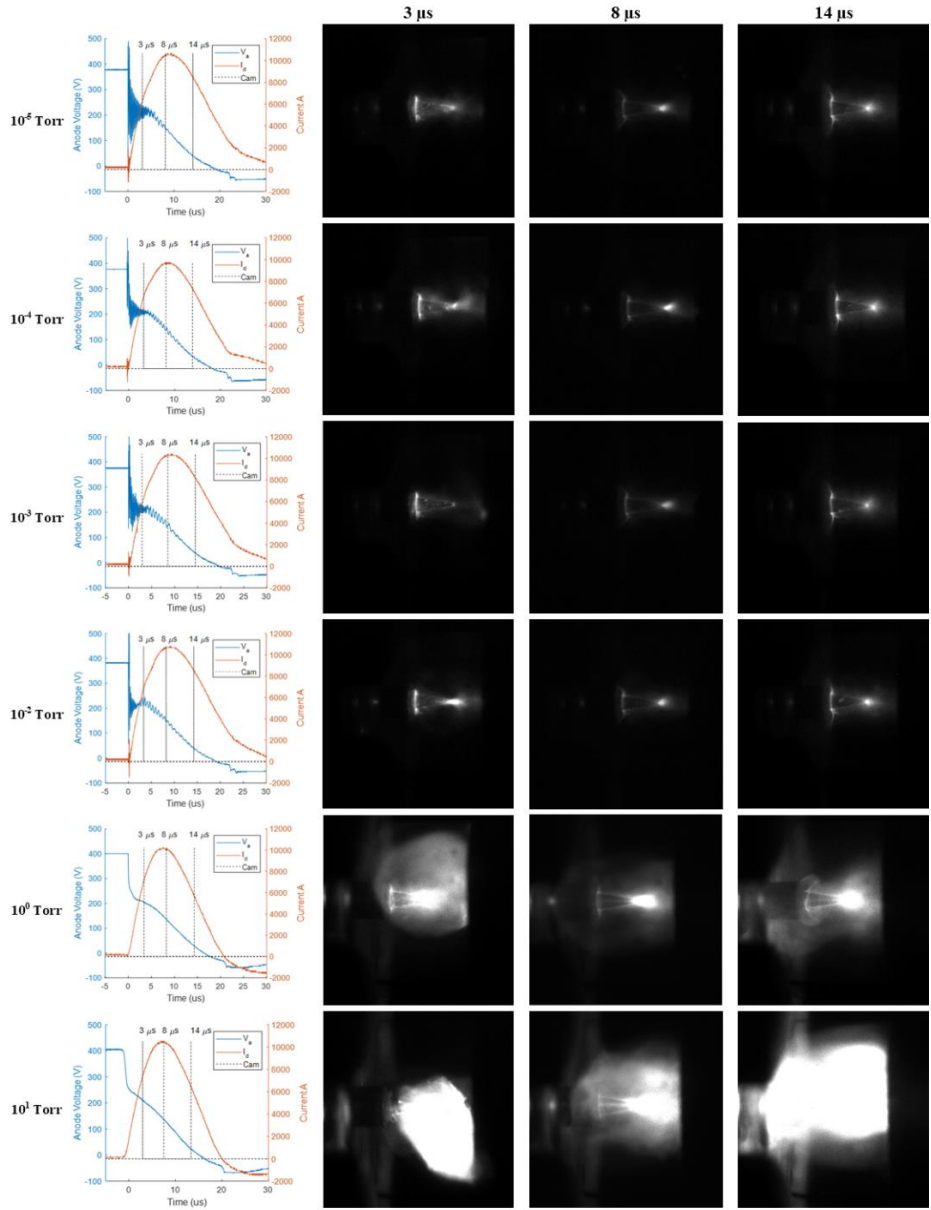

*Figure 5: Images of PPT discharge at different times for different pressures*

Long-exposure images were taken to further analyze the discharge attachment to the cathode at low and high pressures as shown in Figure 6 and Figure 7. These figures depict a high local pressure of 4 Torr and $3.5\times10^{-5}$ Torr, respectively. At higher pressure (4 Torr), one can see that there was a stationary attachment of the discharge to the tip of the cathode, while no signs of conventional cathode spots were observed. Furthermore, the erosion of the tungsten cathode after prolonged operation (13,000 pulses) can be seen in the upper-right image of each figure with the original cathode image as a reference in the lower right of Figure 6. The removal of the cathode tip shown in Figure 6 was most likely due to overheating of the tip as cathode spots were not detected. At a lower pressure ($3.5\times10^{-5}$ Torr), the clear visual presence of cathode spots was detected. As is seen in Figure 7, cathode spots appeared randomly on the cathode surface and there were visually identifiable signs of the cathode consumption after the 13,000-pulse experiment.

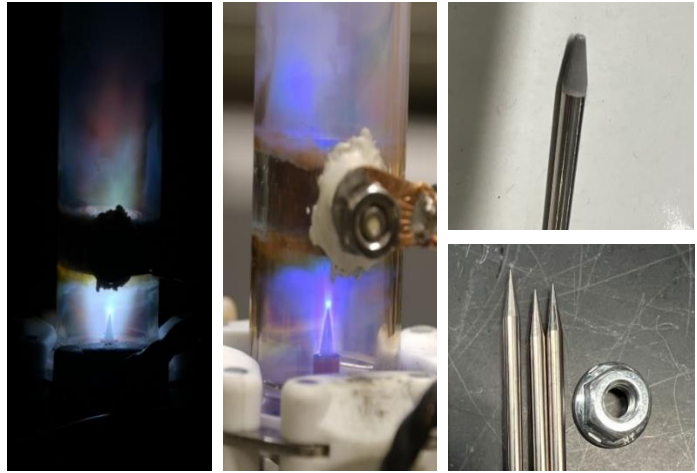

*Figure 6: (Left & Middle) "Spotless" diffuse arc attachment at the tip of the cathode at 4 Torr; (Top Right) Photo of the cathode after 13,000 pulses at 4 Torr; (Bottom Right) Original face of the tungsten cathode with ¼"-20 nut for size comparison*

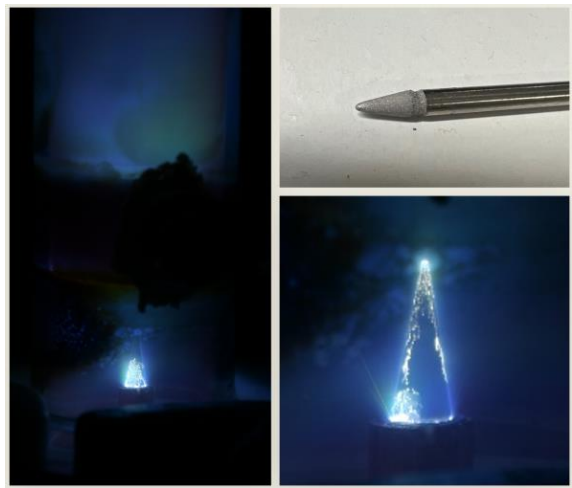

*Figure 7: (Left) Photograph of cathode spots during PPT operation at $3.5\times10^{-5}$ Torr; (Top Right) Visual evidence of the tungsten cathode consumption after 13,000 pulses at $3.5\times10^{-5}$ Torr; (Bottom Right) Zoomed-in view of cathode spots appearing during discharge.*

To quantitatively evaluate the erosion rate of the cathode, we conducted cathode mass loss measurement in long duration tests (over 13,000 pulses). At the higher local pressure (4 Torr), the cathode mass loss after 13,000 pulses was found to be 22.6 mg (the mass loss per pulse was 1.74 µg/pulse), and the erosion rate was $Er = 16.194$ µg/C. For the lower local pressure ($3.5 \times 10^{-5}$ Torr), the cathode mass loss after 13,000 pulses was found to be 77.1 mg (the mass loss per pulse was 5.93 µg/pulse), and the erosion rate was $Er = 40.785$ µg/C. One can see that the erosion rate at lower pressures ($3.5 \times 10^{-5}$ Torr) is consistent with cathodic erosion supported by cathode spots reported in literature [17]. Slightly lower erosion rates observed here could potentially be explained due to the visually observed erosion at the cathode-ceramic interface, as shown in Figure 7. On the other hand, one can see that the erosion rate at higher pressures (4 Torr) was substantially lower which also supports the absence of cathode spots. Additionally, the erosion rate can be further corrected to $Er = 9.93$ µg/C if the cathode tip overheating is eliminated. Note that the prevention of the cathode tip overheating by means of optimizing the cathode's geometry and the system's discharge pulse duration should be the subject of future work.

The experimentally found erosion rates of the tungsten cathode in this PPT configuration are summarized in Table 1.

*Table 1: Erosion rate of the tungsten cathode at different local pressures, including error measurements*

| Pressure (Torr) | Erosion Rate (µg/C) |
|---|---|
| 4.0 | 16.194 +/- 0.022 |
| $3.5 \times 10^{-5}$ | 40.785 +/- 0.0265 |

## V. Conclusion

This work confirms the feasibility of the operation of the PPT considered here at a reduced discharge voltage in the range of $100 - 200 V$ similar to MPD arcs. Furthermore, "spotless" attachment of the discharge column to the cathode surface observed at higher pressures paves the way for the application of this current PPT concept as an electric propulsion device, as the destruction of the cathode can be minimized, and a longer system lifetime can be expected. Future work should be focused on the optimization of the cathode geometry and the discharge's pulse duration to avoid cathode overheating, as well as the designing and testing of the PPT with actual liquid propellant.